# Gas-dynamic acceleration of bodies till the hyper sonic velocity

S. N. Dolya

*Joint Institute for Nuclear Research, Joliot - Curie str. 6, Dubna, Russia, 141980*

**Abstract**

The article considers an opportunity of gas-dynamic acceleration of body from the initial zero velocity till the finite velocity: $V_{fin}$ = 5 km / s. When the gas flow rate of the body pre-acceleration reaches $V_{in}$ = 1 km / s, the body is accelerated at the front of the explosion wave propagating along the coils of the hexogen spiral. This wave accelerates the body and, finally, it reaches the velocity of 5km/s. The accelerated body has mass m = 0.1 kg and diameter $d_{sh}$ = 11.3 mm. Acceleration length is $L_{acc}$ = 6 m. At the slope of the spiral to the horizon equal to $\Theta = 70^0$ the flight range of the body is equal to: $S_{max}$ = 1600 km, and the maximum height of the flight is $H_{max}$ = 1100 km.

**Introduction**

There is a known [1] method of gas-dynamic acceleration of the body in the trunk. The powder disposed in the trunk for a short period of time transfers from the solid to the gaseous state, after that, due to its expansion it pushes the body out from the trunk. The finite velocity of the body in this method of the gas-dynamic acceleration is determined by the density and temperature of the composed gas.

Increasing the density of the gas and its temperature is limited by the durability of the trunk and can not be increased. Modern systems are designed for pressure 3500 atm. and temperature 3000 $C^0$. Further increase of the temperature and pressure beyond these values is not justified because at high temperatures the trunk is damaged and increase of the pressure requires the quality of the trunk to be improved. Further, at high temperatures there is an opportunity of dissociation of molecules of the gas that is related with the loss of the thermal energy of the gas. Therefore, at the present state of the technique development the achieved velocity of the body is the upper limit.

Taking into account the heat dissipation losses due to friction against the walls of the gas trunk and other losses it is possible to accept the limited velocity approximately equal to V ≈ 2 km / s. This velocity is achieved when the body mass is a negligibly small part of the mass of the gunpowder. If you increase the mass of the body, the body velocity will seek to reach the ordinary velocity of the body: $V_{ord} \approx 1$ km / s.

To obtain the hypersonic velocity of the body about $V_{sh} \approx 5$ km / s is impossible by using this method.

In principle, one can accelerate the body by the explosion wave starting with the zero initial velocity of the body. At certain synchronization of the body with the explosion wave it is possible to accelerate the body at non- zero initial velocity. However, this method has a fundamental drawback: it is the small length interaction of the body with the explosion wave.

Let us consider an opportunity of accelerating the body till the hyper sonic velocity formed in the space by the explosion wave propagating synchronically with the accelerated body.

**Selection of basic parameters**

As the explosive substance we have chosen [2] the hexogen cord with a diameter $d_{cord} = 0.5$ cm, which is wound onto the tube with a diameter $2r_0 = 3$ cm. Let the body which is supposed to accelerate, be of a cylindrical shape with mass $m_b = 0.1$ kg, cross-section $S_{tr} = 1$ cm$^2$, and it has the initial velocity $V_{in} = 1$ km / s. The finite velocity of the body is chosen to be equal to $V_{fin} = 5$ km / s and the acceleration length is equal to $L_{acc} = 6$ m.

To obtain this acceleration, we find the following from the equation of motion of the body:

$$m_b dV / dt = F_{axis}, \qquad (1)$$

that the accelerating force must be equal $F_{axis} = 2 * 10^5$ N $= 2 * 10^5$ kg * m/s$^2$. Indeed, dividing this force by body mass $m_b = 0.1$ kg, we find that the body will move with the uniformed acceleration: $a = 2 * 10^6$ m/s$^2$.

The velocity of the body will increase in accordance with the law:

$$V = V_{in} + at, \qquad (2)$$

and from the initial velocity $V_{in} = 1$ km / s till the finite velocity $V_{fin} = 5$ km / s the body will be accelerated during $t_{acc} = 2$ ms.

To implement the impact the permanently acting force onto the body, it is necessary to make an explosion of the hexogen cord wound on the carcass as a double spiral structure. The winding step must permanently grow so that the wave propagation velocity along the spiral would permanently coincide with the velocity of the body.

The dependence of the passed distance (at the uniformed accelerated motion) on the time can be written as follows:

$$L = V_{in}t + at^2/2 . \qquad (3)$$

Solving this equation for the velocity, we find that the dependence of the velocity on the covered distance is expressed as follows :

$$V(L) = (V_{in}^2 + 2aL)^{1/2} . \qquad (4)$$

At the initial parameters, when $L = 0$, the velocity of the body is equal to $V = V_{in} = 1$ km / s. At the finite parameters: $L = L_{acc} = 6$ m, the velocity of the body is equal to: $V = V_{fin} = 5$ km / s.

Knowing the law of the velocity growth depending on the length of the passed distance, it is possible to find the dependence of the winding step on the passed distance.

Assume the velocity of propagation of detonation in hexogen to be equal to $V_{det} \approx 8$ km / s [2]. To have the velocity of the explosion wave $V_{axis}$, propagating along the axis of the spiral, permanently coincide with the velocity of the body $V_b$, the following relation must be satisfied:

$$V_b = V_{axis} \approx V_{det}h/2\pi r_0, \qquad (5)$$

where $V_b$ - velocity of the body, $V_{axis}$ - the velocity of the explosion wave propagating along the axis of the spiral, $r_0$ - radius of the spiral winding, h - the winding step of the spiral.

Detonation propagating along hexogen cord runs around the perimeter of the spiral, along the axis it runs with velocity $V_{axis} \approx V_{det}h/2\pi r_0$. Hence, for the initial velocity of the body, $V_{in} = 1$ km / s, the winding step should be equal to: $h_{in} \approx 2\pi r_0 V_{in}/V_{det} = 1.17$ cm, the finite step of winding of the spiral must be equal to: $h_{fin} \approx 2\pi r_0 V_{fin}/V_{det} = 5.88$ cm. Intermediate values of the winding step of the spiral are given by the following ratio:

$$h(L) = V(L) * V_{det}/2\pi r_0. \qquad (6)$$

We will consider the detonation of the double spiral with a diameter of each of the hexogen cords equal to $d_{cord}$, equal to: $d_{cord} = 0.5$ cm.

Pressure on the detonation front in hexogen reaches $P_{hex} \approx 30$ GPa [2]. Let the body be located at the distance equal to the radius of the spiral winding $z = z_s = 1.5$ cm at the moment when the explosion wave comes up to its back slice.

We will consider the detonation of the long cylindrical hexogen cord. In the spread out of the explosion products the pressure of the explosion wave $P_{wave}$ at the back slice of the body would have been less than the pressure on the front of the wave by the ratio of the square of the radius of the hexogen cord: $(d_{cord}/2)^2 = 0.625$ cm$^2$ to the square of the distance till the back slice of the body (4.5 cm$^2$). It would have been as follows: $P_{wave1} = P_{hex} r^2_{cord} / (r_0^2 + z_s^2) = 0.4$ GPa. Since the construction of detonating chamber is like this, the explosion products leave the area of the explosion only through one quadrant, the pressure at the front of the explosion wave is by 4 times more than in the free space and equal to $P_{wave2} = 1.6$ GPa.

In this case the force acting on the back slice of the body is as follows:

$$F_{axis} = P_{vawe2} * S_{tr} * \cos \varphi, \qquad (7)$$

where we have chosen: $\varphi = 45^0$, $\cos \varphi = 0.7$. Taking into account that two cords are simultaneously detonated, their forces are summed up.

Substituting the numerical values into formula (7) we find that this force is equal to $F_{axis} \approx 2.2 * 10^5$ N.

The light carcass with the hexogen spiral and rings of the special shape, as well as the separable cylinder are located inside the strong trunk with the internal diameter of $D_{bar} = 80$ mm. In this case the pressure onto the inner surface of the trunk at the absence of the substance between the strong trunk and hexogen cord would have been equal to the value: $P_{bar} = P_0 * (d_{cord}/2)^2 / [(D_{bar}/2) - r_0]^2 \approx 3000$ atm., which is the ordinary pressure for the guns [1]. Due to the presence of the rings and separable cylinder the pressure on the inner surface of the strong trunk will be less.

After reaching the velocity V = 2 km / s it is possible to use a quadrifilar spiral. But it is necessary to reduce the diameter of the two cords accordingly (by the root of 2 times). When velocity V = 3 km / s it will be possible to use the six-filar spiral, and etc.

**Synchronization of the body being accelerated and the explosion wave**

As it is shown in [3], the velocity of detonation product spread out of the cylindrical cord in the transverse direction is: $V^\perp = 0.8\ V_{det}$, i.e. In our case this velocity is approximately equal to $V^\perp \approx 6.5$ km / s. Detonation products should reach the back slice of the body at the moment when the back slice is at the distance equal to the radius of the spiral - $l_{in} = r_0 = 1.5$ cm. Then the distance from the back slice of the body to the explosive area of the detonating cord must be $l_{body} \approx 2.12$ cm.

Explosion products fly over this distance during the time $\tau_{shok}$ equal to: $\tau_{shok} = l_{body} / V^\perp = 3.26$ µs. The body moving with the initial velocity $V_{in} = 1$ km / s, must be at this moment at the distance $l_{in} = 3.26$ mm from the start of the acceleration.

**Phase stability principle for the body being accelerated on the front of the explosion wave**

As in particle accelerators on the traveling wave there is the phase stability [4] while accelerating the body at the front of the explosion wave. In the particle accelerators the acceleration phase for particles is chosen in advance, it is called synchronous. For this phase the longitudinal motion of the particles is calculated after that. If any particle is accidentally behind the synchronous phase, it will get into the stronger electric field. Then it will be more accelerated and finally will catch up with the synchronous phase.

If the particle is too fast in its motion for the synchronous phase, then it will get into a smaller electric field, it will be less accelerated, and, finally, the accelerated moving pulse and the synchronous phase will catch up with the particle.

Below we show that in this case while accelerating the body at the front of the explosion wave, the dependence of the force acting for the body on the phase of the pulse at its front, has a declining character.

The force acting on the body, $F_{axis} = P_{wave} * S_{tr} * \cos \varphi$, depends on the pressure $P_{wave} = P_{hex} * r_{cord}^2 / (r_0^2 + z^2)$, where $P_{hex} = 30$ GPa - pressure of the explosion wave (hexogen), $2r_{cord} = 0.5$ cm - diameter of the hexogen cord, $r_0 = 1.5$ cm – the hexogen cord winding radius, z - the distance along the axis from detonating area till the back slice of the body (cm). Projection of the pressure force on the axis of the acceleration is proportional to $\cos \varphi$, which can be represented as follows: $\cos \varphi = z / (r_0^2 + z^2)^{1/2}$.

Thus, the dependence of the force acting on the body along the axis ($F_{axis}$) on the distance along the axis between the detonating area and the back slice of the body, is expressed as follows:

$$F_{axis} \sim z / (r_0^2 + z^2)^{3/2}. \qquad (8)$$

Below we compile a table of the values of this function for three values of the distance between the detonation region and the back slice of the body.

Table 1. Values of the function $z / (r_0^2 + z^2)^{3/2}$

| z, cm | $z/(r_0^2+z^2)^{3/2}$ |
|---|---|
| 1 | 0.17 |
| 1.5 | 0.157 |
| 2 | 0.128 |

From the analysis of this function it is clear that the force acting on the body reduces if the body is faster than the synchronous phase. The explosion force acting on the body increases if the body is behind the synchronous phase. This dependence corresponds to the stable phase of the longitudinal motion according to the phase stability principle.

**Transverse movement of the body**

Let us consider the transverse force available at the front of the explosion wave on the body. In the beginning of the acceleration the velocity of the explosion wave propagating along the axis of the spiral is $V_{in} = 1$ km / s.

The detonation velocity propagating along the spiral is $V_{det} = 8$ km / s.
The perimeter of one of the spiral coils is $\pi d_{spir} \approx 10$ cm, so that the time of detonation of the total cycle is as follows: $\tau_{cyc} \approx \pi d_{spir} / V_{det} = 10$ μs.

Assume that the body moves transversely accelerated but by two orders of the magnitude lower than in the longitudinal direction, i.e., $a = 2 * 10^4$ m/s$^2$. During the time when the detonation runs one third of the total cycle $\tau_{cyc1/3} = 3$ μs, the body is shifted by a distance of $S_{1/3} = a \tau^2_{cyc\ 1/3}/2 = 0.1$ μ. After that the transverse force changes its direction and after the total cycle its action is averaged.

**Ballistics. Aerodynamic resistance**

Now we calculate the motion of the body released at an angle of $\Theta = 70^0$ to the horizon, taking into account the air resistance. The equation of the horizontal motion of the body can be written as follows:

$$m_b\, dV_x / dt = \rho C_x S_{tr} V_x^2 / 2, \qquad (9)$$

where m - mass of the body , $V_x$- horizon velocity, g - 0.01 km/s$^2$ - acceleration due to gravity , $\rho = \rho_0 e^{-z/H0}$ - barometric formula of the change of the atmospheric density in dependence on the height , $\rho_0 = 1.3 * 10^{-3}$ g/cm$^3$ - air density at the Earth surface, $H_0 = 7$ km - the height at which this density decreases by a factor of e.

The aerodynamic coefficient or coefficient of drag resistance is presented as a dimensionless quantity, taking into account the "quality" of the body shape:

$$C_x = F_x / (½) \rho_0 V_x^2 S_{tr}. \qquad (10)$$

The solution of equation ( 9) can be written as follows:

$$V (t) = V_x / [\ 1 + \rho C_x V_x * S_{tr} * t/2m_b\ ]. \qquad (11)$$

In order to calculate the change in velocity of the body in dependence on time, you need to find the aerodynamic coefficient $C_x$.

**Calculation of the aerodynamic coefficient for the air**

We assume that the body has the shape of a cylindrical rod with a conical head. Then at the hit of a nitrogen molecule on the sharp cone, the change of the longitudinal velocity of the molecules is equal to:

$$\Delta V_x = V_x * \Theta_t^2 / 2, \qquad (12)$$

where $\Theta_t$ –the cone angle at the vertex. Gas molecules transfer the momentum to the body:

$$p = mV = \rho V_x S_{tr} t * \Delta V_x. \qquad (13)$$

The change in the momentum per unit of the time is the force which is called the force of the frontal slow down,

$$F_{x1} = (½) \rho V_x S_{tr} * V_x * \Theta_t^2. \qquad (14)$$

Dividing $F_{x1}$ by $(½) \rho V^2_x S_{tr}$, we obtain the drag coefficient for the sharp cone at the mirror reflection of the molecules from the cone, (Newton's formula):

$$C_{x\ air} = \Theta_t^2. \qquad (15)$$

Our consideration corresponds to hypersonic velocity; we can neglect the effects that occur at the velocity close to the sonic velocity in the unperturbed medium. Let the length of the conical part of the body be as follows: $l_{cone} = 65$ mm and the diameter of the body $d_b = 11.3$ mm. This means that the angle at the vertex of the cone is: $\Theta_t = 0.173$ and $C_{x\ air} = 0.03$.

Substituting this value in the formula for the $C_x$ loss of the longitudinal velocity in dependence on time (11), we find that the decrease in the vertical velocity of the body in the first second of the flight is of the order of 10%.

The flight range of the body is $S_{max} = 2V^2_0 * \sin\Theta * \cos\Theta / g = 1600$ km, the maximum lifting height of the body is $Y = V^2_0 * \sin^2\Theta/2g = 1100$ km.

Fig. 1 shows a diagram of the device.

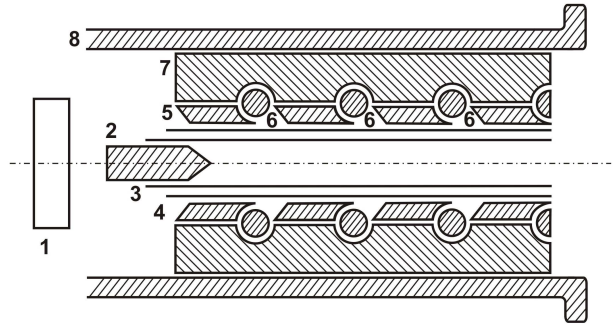

Fig.1. 1 – the gun, 2 - the body, 3 – the light trunk, 4 - light carcass,
5 - special shaped rings, 6 - hexogen cord, 7 - separable cylinder,
8 - solid trunk.

The operation of the device is as follows. In the gun (1), the acceleration of the body (2) of a cylindrical shape with mass $m_b = 0.1$ kg, and the cross-section of $S_{tr} = 1$ cm$^2$ is performed by using the ordinary gas - dynamic method. The body is centered relatively the spiral with a light trunk (3). The body moves in the light carcass (4), where the special shaped rings (5) are embedded. Synchronously with the start of the acceleration we produce the explosion of the hexogen cord (6), which is placed in the gaps between the rings. Outside the cord is surrounded with the separable cylinder (7). The rings must be rigidly fixed with the separable cylinder. Figure 1 does not show fastening of the rings to the cylinder. All the assembly is placed inside the solid trunk (8). Detonation propagates along the cord and creates the explosion wave running synchronously with the body.

If to increase the mass of the body by an order of the magnitude till the value of $m_{b1} = 1$ kg, then the length of the acceleration (at the same parameters) will increase by the order of magnitude and will be equal to the value of $L_{acc1} = 60$ m. Such an accelerator may be arranged only horizontally, while at the same time the body must have asymmetry which creates the lifting force.

Suppose that the lifting coefficient is, $C_y = 0.2$. In this case the drag coefficient must be the same: $C_x = 0.2$.

Thus, the equation of the vertical motion can be written as follows:

$$m_{b1} dV_y / dt = C_y \rho_0 V_x^2 * S_{tr} / 2, \qquad (16)$$

where $C_y$ is the aerodynamic lifting force coefficient, $\rho_0 = 1.3 * 10^{-3}$ g/cm$^3$ - the air density on the surface of the Earth, $V_x = 5$ km / s - the horizontal velocity of the body, $S_{tr}$ is the cross-section of the body.

Solving approximately equation (16), we obtain:

$$V_y = C_y \rho V_x^2 S_{tr} t / 2 m_{b1}. \qquad (17)$$

Integrating again, we obtain an expression for the lifting height of the body:

$$Y_1 = C_y \rho V_x^2 S_{tr} t^2 / 4 m_{b1}. \qquad (18)$$

Solving the equations of horizontal and vertical movements we can obtain the time dependence of the flight parameters of the body, which we present in Table 2. The first column gives the time of the flight, the second - the horizontal velocity, in the third – there is the vertical velocity, the fourth column shows the height of the lifting of the body.

Table 2. Flight parameters for the case of $C_x$, $C_y = 0.2$.

| t, s | $V_x$, km/s | $V_y$, km/s | Y, km |
|---|---|---|---|
| 0 | 5 | 0 | 0 |
| 2 | 4.43 | 0.65 | 0.65 |
| 5 | 3.8 | 1.37 | 2.7 |
| 10 | 3 | 2.3 | 8.5 |
| 20 | 2.68 | 2.66 | 20 |

The time of the body lifting till the maximum height in this case is as follows: $\tau_{max} = V_y / g = 266$ s. The flight range of the body is $S_{max1} = V_x * 2\tau_{max} = 1400$ km, the maximum body lifting height is $Y_1 = V_y^2 / 2g = 350$ km. Changing the shape of the cone in the head part of the body, it will be possible to obtain different trajectories.

**Conclusion**

These flight parameters may be of interest for a number of applications, in particular, for the removal of the space garbage at the low orbits.